\documentclass{aastex}
\usepackage{apjfonts}
\usepackage{myemulateapj5}
\newcommand{\timav}{\langle \dot M \rangle}

\shorttitle{Faint CVs in Quiescence: GC \& Field Surveys}

\begin{document}

\submitted{To Appear in Astrophysical Journal Letters}

\title{Faint Cataclysmic Variables in Quiescence: Globular Cluster
and Field Surveys}

\author{Dean M. Townsley}
\affil{Department of Physics\\
Broida Hall, University of California, Santa Barbara, CA 93106;
townsley@physics.uscb.edu}
 
\author{Lars Bildsten} 
\affil{Institute for Theoretical Physics and Department of Physics\\
Kohn Hall, University of California, Santa Barbara, CA 93106;
bildsten@itp.ucsb.edu}

\begin{abstract}

 Current evolutionary models imply that most Cataclysmic Variables (CVs) have
$P_{\rm orb}<2 $ hours and are Dwarf Nova (DN) systems that are quiescent
most of the time. Observations of nearby quiescent DN find that the UV
spectrum is dominated by the hot white dwarf (WD), indicating that it
provides a significant fraction of the optical light in addition to the
quiescent disk and main sequence companion.  Hence, identifying a faint,
quiescent CV in either the field or a globular cluster (GC) from broadband
colors depends on our ability to predict the WD contribution in quiescence.
We are undertaking a theoretical study of the compressional heating of WDs,
extending down to very low time averaged accretion rates, $\timav\sim
10^{-11}M_\odot \ {\rm yr^{-1}}$, which allows us to self-consistently find
the $T_{\rm eff}$ of the WD. We demonstrate here that most of the
compressional heating occurs in the freshly accreted envelope and that the WD
core temperature reaches a fixed value on a timescale less than typical
evolutionary times. Since nuclear burning is unstable at these $\timav$'s, we
have incorporated the recurrent heating and cooling of the WD core throughout
the classical novae limit cycle in order to find the $T_{\rm eff}$-$\timav$
relations. Comparing to observations of field DN confirms the
$\timav$-$P_{\rm orb}$ relation of disrupted magnetic braking.  We also
predict broad-band colors of a quiescent CV as a function of $\timav$ and
companion mass and show that this leads to the identification of what may be
many CVs in deep HST images of GCs.

\end{abstract}


\section{Introduction} 

  Dwarf Novae (DN) systems contain a white dwarf (WD) accreting matter at
time-averaged rates $\timav<10^{-9}M_\odot \ {\rm yr}^{-1}$ from a low-mass
($<0.5M_\odot$ typically) stellar companion (see Osaki 1996 for an overview).
At these $\timav$'s, the accretion disk is subject to a thermal instability
which causes it to rapidly transfer matter onto the WD (at $\dot M \gg
\timav$) for a few days to a week once every month to year.  The orbital
periods of these binaries are usually less than 2 hours (below the period
gap), but there are also DN above the period gap, $>$ 3 hours (see Shafter
1992).  The $\dot M$ onto the WD is often low enough between outbursts that
the UV emission is dominated by the internal luminosity of the WD.
Indeed recent spectroscopy has resolved the WD's contribution to the
quiescent light and found effective temperatures $T_{\rm eff}\sim
10,000-40,000 \ {\rm K}$ (see Sion 1999).

 The measured internal WD luminosity is larger than expected from an isolated
WD of similar age ($\approx$ Gyr), indicating that it has been heated by
accretion (Sion 1985).  Compressional heating (i.e. internal gravitational
energy release) appears to be the main driver for this re-heating (Sion
1995).  Sion's (1995) estimate for internal gravitational energy
release within the WD (of mass $M$ and radius $R$) was $L\approx 0.15
GM\timav/R$.  However, we show in \S \ref{sec:physics} that the energy
release actually depends on the thermal state of the WD interior and that the
dominant energy release is in the accreted outer envelope, giving $L\approx
3kT_c \timav/\mu m_p$, where $\mu\approx 0.6$ is the mean molecular weight of
the accreted material, $T_c$ is the WD core temperature, $m_p$ is the baryon
mass, and $k$ is Boltzmann's constant.

The theoretical challenge that we address in \S \ref{sec:Tcmethod} is how to
calculate $T_c$ as a function of $\timav$, and thus find $T_{\rm eff}$.
Because of unstable nuclear burning and the resulting classical novae cycle,
the H/He envelope mass changes with time, allowing the core to cool at low
accumulated masses and be heated prior to unstable ignition. We use nova
ignition to determine the maximum mass of the overlying freshly accreted
shell, and find the steady-state (i.e.  cooling equals heating throughout the
classical novae cycle) core temperature, $T_c$, as a function of
$\timav$ and $M$.

We compare our calculations to HST/STIS
observations and infer $\timav$ on the timescale of $10^6$ years.  We find
that DN above the period gap have $\timav\approx 10^{-9}M_\odot \ {\rm
yr^{-1}}$, while those below have $\timav\approx 10^{-10} M_\odot \ {\rm
yr^{-1}}$, consistent with that expected from traditional CV evolution (e.g.
Howell, Nelson, \& Rappaport 2001), even those that involve some
``hibernation'' (Shara et al.\ 1986; Kolb et al.\ 2001).  The result is more
surprising if the much weaker magnetic braking laws of Andronov,
Pinsonneault, \& Sills (2001) are correct.  We also predict the minimum light
($M_V$) of CVs in quiescence for a range of $\timav$, WD mass, and companion
mass.  This assists the search for the predicted large population of CVs with
very low mass companions ($<0.1M_\odot$) that are near, or past, the period
minimum (Howell, Rappaport, \& Politano 1997).  Observations already show
that the WD fixes the quiescent colors of these CVs and our calculations are
useful for CV surveys in the field (e.g. 2DF, SDSS, see Marsh et al.\ 2001 and
Szkody et al.\ 2002) and globular clusters.

\section{Compressional Heating: Core--Envelope Contrast}
\label{sec:physics}

Compressional heating is the energy released by fluid elements as they are
compressed by further accretion.  The important feature of this heating
mechanism is that the heat is released in the WD \textit{interior},
and thus is radiated on a timescale which is longer than the time between
DN outbursts.  Contrast this to the gravitational potential
energy released by the infalling matter, ($GMm_p/R$ per baryon) which is
deposited at, or near, the photosphere and is rapidly radiated away.  Such
infall energy does not get taken into the star because in the upper
atmosphere (where $T\ll T_c$) the time it takes the fluid to move inward is
much longer (by at least $T_c/T$) than the time it takes
for heat to escape.  This means infall energy is not important for
setting the internal thermal state of the WD; it simply has no influence
there.  Also, once accretion has diminished for longer than the
time to radiate the infall energy away (such as in DN quiescence), it is no
longer relevant to the observed luminosity.

An additional energy source in the WD interior is slow nuclear
burning near the base of the accreted H/He layer.  This is significant
when the accreted layer becomes thick,  eventually becoming thermally
unstable and leading to a Classical Nova.  The nova energy is assumed to be
radiated away in the explosion, but we have found that slow burning before
that point contributes an amount of energy to the WD interior comparable to
the energy released by compression.  Our calculations take account of both
compressional heating and slow nuclear burning to determine the WD's thermal
state.

  We now sketch how compressional heating is included in our stellar model,
demonstrate that heating in the envelope dominates that in the
core and estimate its magnitude.  This is a  discussion specific to DN with
low $\timav$, and the reader should consult Nomoto (1982) for a complete
account.  The simplest estimate of the compressional energy release is the
gravitational energy liberated as a fluid element moves down in the WD
gravitational field, $g=GM/R^2$.  In the non-degenerate outer atmosphere, a
fluid element moves a distance of order the scale height, $h=kT/\mu
m_p g$, in the time it takes to replace it by accretion, giving $L\sim \timav
g h\sim \timav k T/\mu m_p$.  This exhibits the correct scaling, notably the
dependence on $\mu$ which is a contrasting parameter between the accreted
H/He envelope and the C/O core.

To calculate the actual heat release we consider the local heat equation
\begin{equation}
\label{eq:heateq}
T\frac{ds}{dt}=T\frac{\partial s}{\partial
t}+T\vec v \cdot \nabla s= -\frac{\partial L}{\partial M_r} +\epsilon_N,
\end{equation} 
where $\epsilon_N$ is the nuclear burning rate, $s$ is the entropy, and
$\vec
v=-\timav \hat r/4\pi r^2 \rho$ is the slow downward advection speed from
accretion.  The entropy profile is fixed by the temperature gradient needed
to carry the luminosity outward and thus we simultaneously solve equation
(\ref{eq:heateq}) with the heat transport equation, using opacities and
conductivities from Iglesias \& Rogers (1996) and Itoh et al.\ (1983) to find
the thermal structure of the accreted envelope and the outer edge of the C/O
core.  For an analytic understanding, we neglect nuclear burning and
$\partial/\partial t$
and use hydrostatic balance to recast equation
(\ref{eq:heateq}) into
\begin{equation} L=-\timav \int_0^P T \frac{\partial s}{\partial P} dP\ .
\end{equation}
Entropy decreases inward (i.e. the envelope and core are not convective),
so this is an outward $L$.  In the non-degenerate envelope,
where $s=k\ln(T^{3/2}/\rho)/\mu m_p$, one more approximation is necessary to
obtain an analytic form.  For an atmosphere in which $L$ is constant with
depth, the envelope satisfies $T^{8.5}\propto P^2$.  Though $L$ is not
constant here, we use this to get an estimate, integrating down to the
isothermal core and finding $L\approx 3kT_c \timav/\mu m_p$.

 Now consider the degenerate C/O core. For the
$\timav$'s and typical $M=0.6 M_\odot$ WD of interest here, the
entropy is in the liquid ions at $T_c\approx 10^7 \ {\rm K}$. The time
it takes to transport heat through the interior is $\sim 10^7 {\rm
yr}\ll M/\timav$, so the core is isothermal and any compression is far
from adiabatic.\footnote{This is in contrast to the rapid accretion
rates $\timav \gg 10^{-8} M_\odot \ {\rm yr}^{-1}$ considered for more
massive Type Ia progenitors, where the interior undergoes nearly
adiabatic compression (see Bravo et al.\ 1996).}  Due to uncertainty
from the classical novae cycle, we don't know whether the C/O core is
secularly increasing in mass, but if it were, almost all of the work
of compression goes into increasing the electron Fermi energy.  The
integrated heat release in the core would then be $L\approx 15 kT_c\timav/\mu_i
m_p$ (Nomoto 1982) for a $0.6M_\odot$ C/O WD, where $\mu_i\approx 14$
is the ion mean molecular weight.

  Due to the mean molecular weight contrast between the accreted envelope and
the core, the energy release in the core is about a factor of five smaller
than that in the envelope.  Thus, for a given amount of compression of the
star, the entropy drop for material in the accreted layer is much larger than
for material in the core.  {\it Despite its comparatively small mass, the
accreted layer is the main source of compressional heating.}

\section{ Finding the Equilibrium Core Temperature}
\label{sec:Tcmethod}

 For this initial study, we dropped the time-dependent term in
equation (\ref{eq:heateq}), and presumed that the C/O core mass was
constant throughout the classical nova cycle, thus only accounting for
the compressional heating and $\epsilon_N$ in the accreted
layer. This method improves on that of Iben et al.\ (1992) by allowing
the accreted envelope mass to change through the $10^5$ year classical
nova cycle. Early in the cycle, the mass of the accreted layer
is small, compressional heating is small, and the WD cools. Later in
the cycle, the accreted layer becomes thick enough that compressional
heating along with slow hydrogen burning releases a sufficient amount
of energy to heat the core.  As the WD has a large heat capacity,
reaching the equilibrium $T_c$ where the heat exchanged between the
envelope and core averages to zero over a single classical nova cycle
takes $\approx 10^8$ years. Since this time is shorter than
the time over which $\timav$ changes due to changing orbital period, we
construct such equilibrium accretors for a given $M$ and $\timav$.

\begin{figure}[t]
\plotone{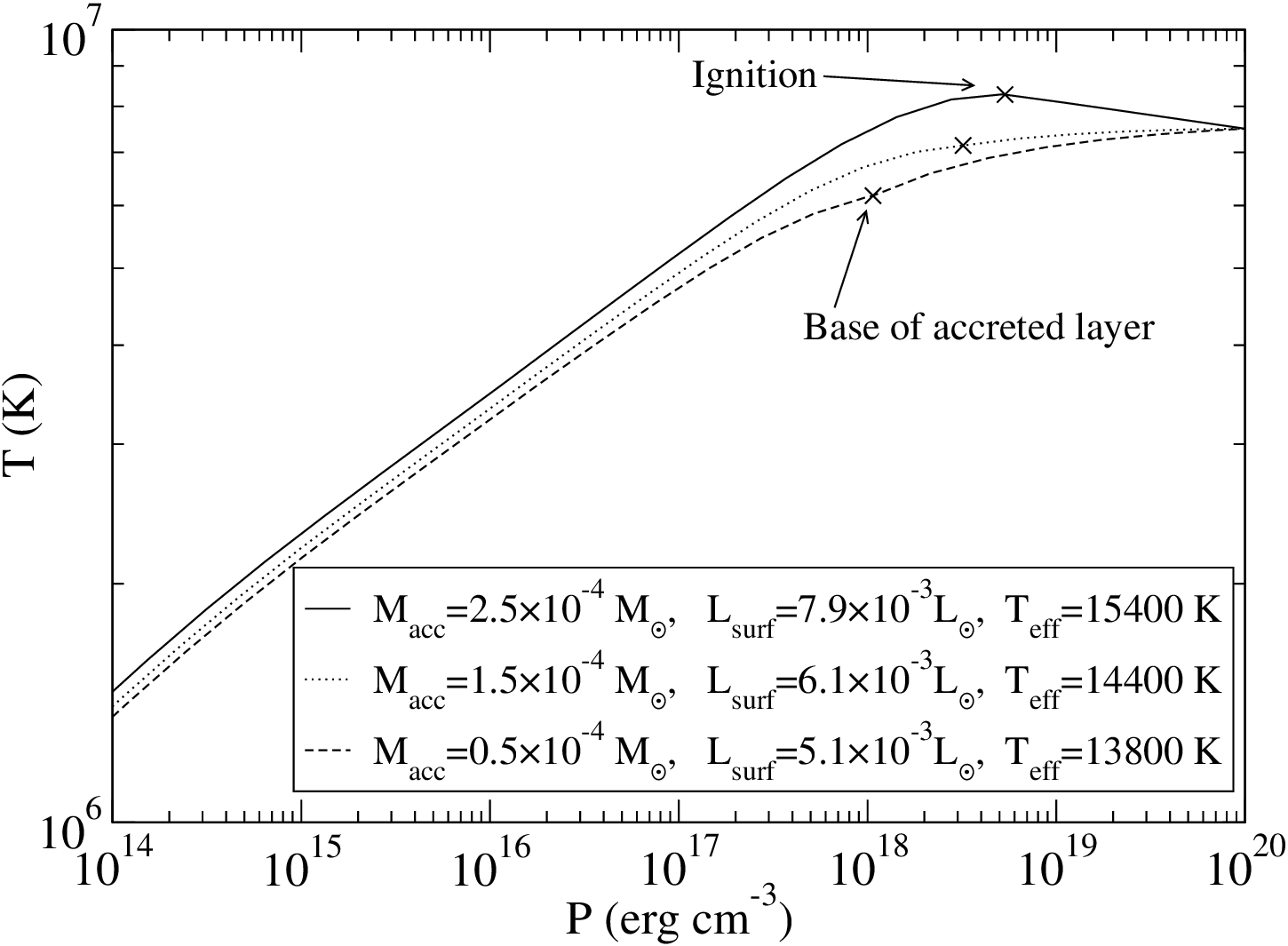}
\caption{\label{fig:T-P}
The Hydrogen/Helium envelope
and outer core in temperature and pressure for three different values
of accumulated mass, $M_{\rm acc}/10^{-4}M_\odot = 0.5$, 1.5,
and 2.5 for $M=0.6M_\odot$ and
$\timav=10^{-10}M_\odot\ {\rm yr}^{-1}$.  The external surface luminosity of the
WD and the corresponding $T_{\rm eff}$ is also listed.  The part of
the star off to the right of the figure ($P>10^{20}$ erg cm$^{-3}$)
is the isothermal inner core.}
\end{figure}

 To do this construction, we first fix $T_c$ at the outer edge of the C/O
core at a pressure high enough so that the changing accumulated mass has
little direct effect.  With a radiative-zero outer boundary condition, we
integrate our structure equations with equation (\ref{eq:heateq}) to
find the thermal state for an $\timav$ and accreted layer mass.  See Figure
\ref{fig:T-P} for examples of the resulting $T$-$P$ relations.  We then
evaluate the luminosity across the chosen location (the right edge of the
plot in Figure \ref{fig:T-P}) for different accreted layer masses, up to the
unstable ignition which is found by comparing the $T$ and $\rho$ at the base
of the accreted layer with analytic ignition curves (Fujimoto 1982).

\begin{figure}[t]
\plotone{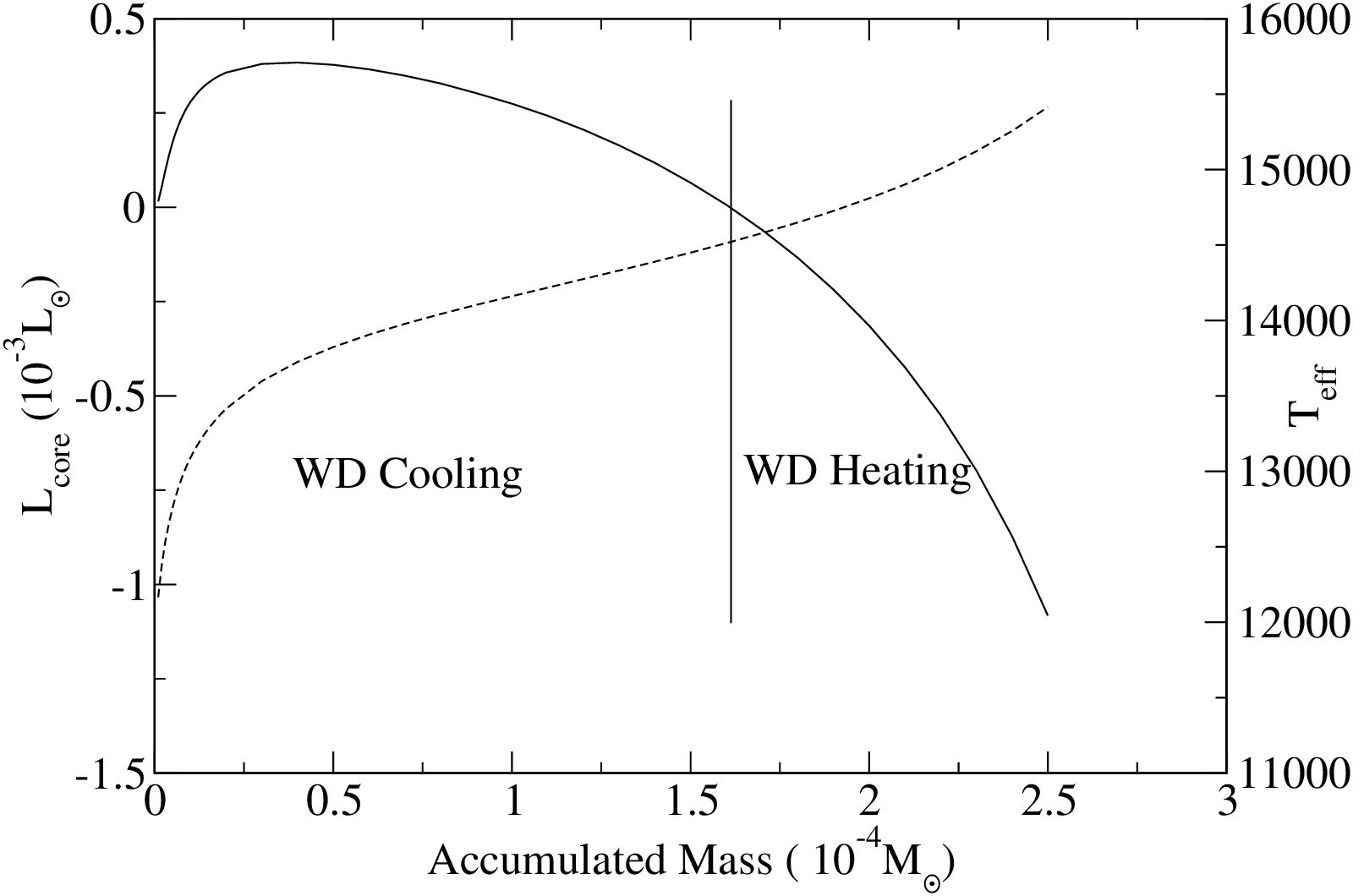}
\caption{\label{fig:loop} \label{fig:Teff}
The solid line is the luminosity
at the outer edge of the core and the dashed is the surface effective
temperature of the WD, both as a function of accumulated mass up to the
classical nova ignition for
the equilibrium model of Figure \ref{fig:T-P}, $M=0.6M_\odot$,
$\timav=10^{-10} M_\odot$ yr$^{-1}$.
Positive luminosity is outgoing and epochs of core
cooling and heating are indicated.
}
\end{figure}

We vary $T_c$ to find an equilibrium model, where the ``core luminosity''
($L_{\rm core}$) averages to zero over the classical nova cycle as shown in
Figure \ref{fig:loop}. The quiescent $T_{\rm eff}$ for the same cycle is
also shown in Figure \ref{fig:Teff}. At the nova outburst we assume that the
accreted shell is expelled, and that, due to the rapidity of this event, it
does not appreciably heat the WD.  The resulting equilibrium core
temperatures for $\timav=10^{-10}M_\odot\ {\rm yr}^{-1}$ are $T_c/10^6{\rm
K}=9, 7.5$ and $8.5$ for $M=0.4, 0.6$ and $1.0M_\odot$.  The $0.4M_\odot$
star is hotter than the $0.6M_\odot$ star because it has a larger maximum
accumulated mass that leads to a longer period of core heating. For a
$0.6M_\odot$ WD, the core temperatures are $T_c/10^6{\rm K}=4,5.3,12.2$ and
18.0 for $\timav/M_\odot \ {\rm yr^{-1}}=10^{-11},3.2\times 10^{-11},
4.2\times 10^{-10}$ and $10^{-9}$.

The $T_{\rm eff}$ during the classical nova cycle varies over a relatively
narrow range that allows us to compare to observations.
For field CVs, the
large set of STIS observations by Szkody et al.\ (2001) and previous
observations (Urban et al.\ 2000) provide
spectra of quiescent WDs in DN.
These measurements are made during deep quiescence when the
accretion luminosity is negligible and are long enough
after the outbursts that other emission mechanisms (e.g. Pringle's
(1988) suggestion of radiative illumination of the WD) have faded. 
The observed $T_{\rm eff}$'s thus measure the heat directly
from the WD interior.
This comparison to observations
indicates that below the period gap, $\timav\approx
10^{-10}M_\odot\ \rm yr^{-1}$ and the WD masses are in the range
$0.6$-$1.0M_\odot$.
 This agrees with the expectation from Kolb \&
Baraffe (1999), who find $\timav\approx 5\times 10^{-11} M_\odot \
{\rm yr}^{-1}$ at an orbital period of 2 hours presuming angular
momentum losses from gravitational waves alone. Above the period gap, the
$T_{\rm eff}$ is higher, and we estimate $\timav\approx
10^{-9}M_\odot\ \rm yr^{-1}$.
For a graphical comparison, see Townsley \& Bildsten (2001).
This general agreement with data from field WDs
in which the internal luminosity is directly visible gives us confidence that
our calculations can be applied to other quiescent DN systems.

 We predict that a $0.6M_\odot$ WD above the gap has
$T_c=1.8\times 10^7$ K and, if in equilibrium below the
gap, $T_c=7.5\times 10^6$ K.
However, if the WD does not have time to cool as it traverses the gap, it
will be hotter than our calculation implies.
We estimate this cooling time from the current  WD cooling
law (e.g. Chabrier et al.\ 2000),
$L_{\rm cool} \approx 10^{-2}L_\odot (T_c/1.8\times
10^7 {\rm K})^{2.5}$, along with the heat capacity of the core,
$M\,3k_B/\mu_im_p$,
giving $\Delta t\approx 0.5$ Gyr. Since this is comparable to the
estimated time spent in the gap (Howell et al.\ 2001), our equilibrium
assumption below the gap is likely safe. However, note that about 0.2
Gyrs after accretion halts, the WD will enter the ZZ Ceti instability
strip!

\section{Application to Globular Cluster Populations}

 Due to the high frequency of stellar interactions in Globular Clusters (GCs),
an abundant population of CVs is expected to be found there, especially at
low $\timav$.  CVs in GCs are commonly searched for via the presence of
hydrogen emission lines or X-ray emission (as recent Chandra observations
have found; Grindlay et al.\ 2001a, Grindlay et al.\ 2001b), and this method is
fruitful.
We show
that these systems (as well as CVs crossing the period gap or those
``hibernating'' post-novae, Shara et al.\ 1986) can also be identified by their
position in a color-magnitude diagram (CMD). By using our theory of the
thermal state of the WD, it is possible to predict the broadband colors of
quiescent CVs.

\begin{figure}
\plotone{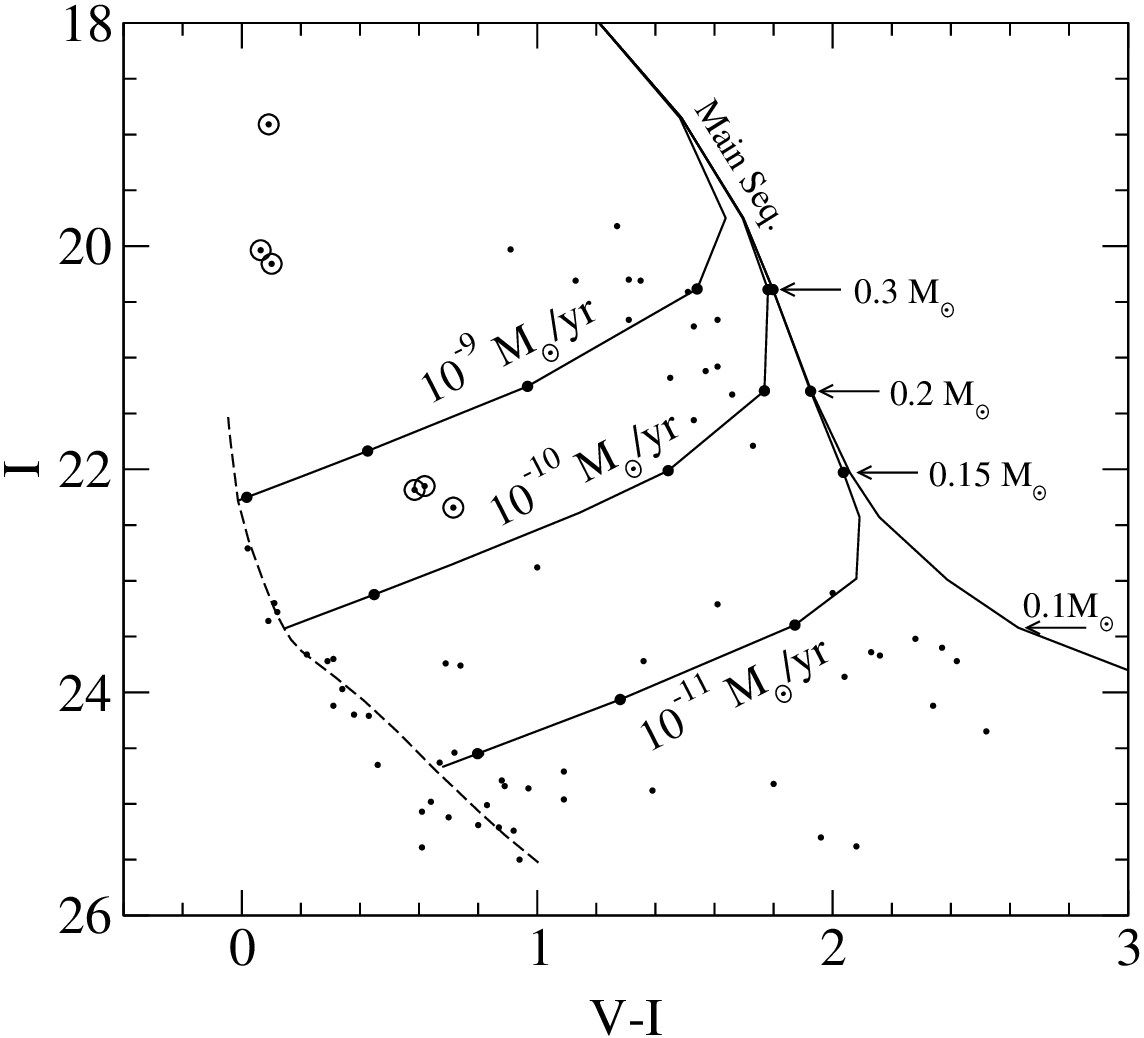}
\caption{\label{fig:gc} Color-Magnitude Diagram of NGC 6397.  The dots
are plausible cluster members that are below the MS (HST observations by
King et al.\ 1998) and the $\odot$'s are the ``Non-Flickerers'' from
Taylor et al.\ (2001). The MS line is from Baraffe et al.\ (1997) for the
cluster [M/H] of $-1.5$ and an age of 10 Gyr. The
dashed line is from Bergeron et al.\ (1995) for DA WDs with $\log
g=8$. The lines connecting the MS to the WD sequence are our current
calculations of the WD+MS binary at the specified $\timav$. The
highest $T_{\rm eff}$ during the classical nova cycle has been used
for the WD in each case (see Figure \ref{fig:Teff}). No disk has been
included. All curves have been put at the distance and reddening of the
cluster, $(m-M)_I=12.05$ and $E(V-I)=0.288$.}
\end{figure}

 An excellent example is NGC 6397 (King et al.\ 1998; Taylor et
al.\ 2001). Figure \ref{fig:gc} shows a CMD of NGC 6397 with our
initial results.
The data points are objects which meet the proper-motion criteria for cluster
membership and which are below the MS.
 The lines were produced by superposing a WD with
the maximum $T_{\rm eff}$ for the indicated $\timav$ with a MS
star.
Due to uncertainty in the theory of quiescent disks (Menou 2000), no disk
contribution has been added.  Note, however, that a constant $T\sim 5000$ K
disk like those indicated in eclipse maps (Wood \& Crawford 1986) would have
a $V-I$ color of 1.24 including the cluster reddening.
Except for near the WD cooling line (dashed curve), where the WD
dominates, the $I$ magnitude is set by the MS
companion.  The large dots along the $10^{-9}M_\odot \ {\rm yr}^{-1} $
and $10^{-10}M_\odot \ {\rm yr}^{-1}$ lines indicate where the MS
companion is 0.3, 0.2, 0.15 and 0.1 $M_\odot$, and two additional
points at 0.09 and 0.085 $M_\odot$ are indicated on the
$10^{-11}M_\odot \ {\rm yr}^{-1}$ line. This immediately provides a
number of candidate systems (namely, data in this part
of the CMD).

 The circled points are the ``non-flickerers'' (Cool et al.\ 1998)
recently reported by Taylor et al.\ (2001). The three at $I\approx
22.25$ are very strong H$\alpha$ absorbers (consistent with a DA WD)
and were not detected by Chandra (Grindlay et al.\ 2001b). These
authors had discussed these systems as possible helium WDs with
millisecond pulsar companions, though, given our work, we would claim
that these are hot WDs with $\approx 0.15 M_\odot$ MS companions.
In addition, the population of data points in this diagram with respect
to our theoretical curves will eventually constrain CV evolutionary
scenarios.  If we assume many of the data points are CVs, we already
see that most systems with high $\timav$ have $0.15-0.3 M_\odot$
companions. If confirmed as members of the cluster, the
stars below the $\timav=10^{-11}M_\odot \ {\rm yr}^{-1}$ line could well be
the long-sought post-turnaround systems with $\timav=10^{-12}M_\odot \ {\rm
yr}^{-1}$ and companion masses $< 0.09 M_\odot$ (Howell et al.\ 1997).

\section{Conclusions and Discussion}

We have evaluated the action of compressional heating on accreting WD
interiors and shown that most of the compressional energy release takes place
in the accreted envelope, and is thermally communicated to the core.
The
maximum envelope mass is set by the unstable nuclear burning that
causes a classical nova runaway and most likely expels the accreted
mass. We have constructed equilibrium accretors which have constant
core temperatures such that the heat lost from the core when the
envelope is thin (i.e. right after the classical nova) is balanced by
that regained when the envelope is thick. This equilibrium determines
the $T_{\rm eff}$ of the WD throughout the classical nova cycle.  Our
models agree with the observations of dwarf novae in deep quiescence
and imply $\timav\approx10^{-10}M_\odot$ yr$^{-1}$ just below the
period gap and $\timav\approx10^{-9}M_\odot$ yr$^{-1}$ just above the
period gap for WD masses in the range $0.6$--$1.0M_\odot$.

 Our $T_{\rm eff}$ calculations provide a prediction of the  colors of
quiescent DN.  Using MS stellar models, we have predicted where a DN should
appear in a color-magnitude diagram as a function of $\timav$ and the mass of
its companion.  Many unidentified objects appear in the relevant regions of
the detailed CMDs which have been obtained for globular clusters by HST.  The
number of such systems in the field will increase due to upcoming
surveys (such as SDSS and 2DF, see Marsh et al.\ 2001), and will push to lower
$\timav$ systems.


 Though our initial efforts have met with apparent success, there is still
much to be done. We need to vary the metallicity of the accreted
material, lowering to values appropriate for globular cluster science.
This could change our results at large $\timav$, but at low $\timav$'s, the
ignition mass is set by $pp$ burning and will likely not change too much.
We also need to relax our initial assumptions, e.g. by
including WD excavation or accretion and accounting for thermal evolution of
the WD.

   The internal thermal state of the WD has been a longstanding uncertainty in
classical nova work, as has the question of how much mass is ejected in the
explosion (Gehrz et al.\ 1998).
Our work provides the first calculation of the internal thermal state of a WD
undergoing classical novae, and will eventually lead to self
consistent calculations for ignition masses, including variations of the
metallicity.
This will be an improvement on previous work (e.g. Prialnik \& Kovetz 1995)
which treated $T_c$ and $\timav$ as two independent parameters.

\acknowledgements 

 We thank Paula Szkody for helpful comments as referee and for sharing the
most recent HST observations, and Ivan King and Jeno Sokoloski for comments.
This research was supported by NASA via grant NAG 5-8658 and by the NSF under
Grants PHY99-07949 and AST01-96422.  L. B. is a Cottrell Scholar of the
Research Corporation and D. T. is an NSF Graduate Fellow.

\clearpage


\begin{references}


\noindent 
Andronov, N., Pinsonneault, M., \& Sills, A. 2001, ApJ, submitted
(\mbox{astro-ph/0104265})

\noindent 
Baraffe, I., Chabrier, G., Allard, F., \& Hauschildt, P. H. 1997, A\&A,
327, 1054

\noindent 
Bergeron, P., Wesemael, F., \& Beuchamp, A. 1995, ApJ, 449, 258 

\noindent 
Bravo, E., Tornambe, A., Dominguez, I., \& Isern, J.  1996, A\&A, 306, 811

\noindent 
Chabrier, G., Brassard, P., Fontaine, G. \& Saumon, D. 2000, ApJ, 543,
 216 

\noindent 
Cool, A. M., et al.\ 1998, ApJ, 508, L75 

\noindent 
Fujimoto, F. Y. 1982, ApJ, 257, 767

\noindent 
Gehrz, R. D., Truran, J. W., Williams, R. E., \& Starrfield, S. 1998,
PASP, 110, 3 


\noindent 
Grindlay, J. E., Heinke, C., Edmonds, P. D., \& Murray, S. S. 2001a,
Science, 292, 2290

\noindent 
Grindlay, J. E., Heinke, C., Edmonds, P. D., Murray, S. S., \& Cool,
A. M. 2001b, ApJ, 563, L00

\noindent 
Howell, S. B., Nelson, L. A., \& Rappaport, S. 2001, ApJ, 550, 897

\noindent 
Howell, S. B., Rappaport, S., \& Politano, M. 1997, MNRAS, 287, 929 

\noindent 
Iben, I., Fujimoto, F. Y., Masayuki Y., \& MacDonald, J. 1992, ApJ, 384, 580

\noindent 
Iglesias, C. A., \& Rogers, F. J. 1996, ApJ, 464, 943

\noindent 
Itoh, N., Mitake, S., Iyetomi, H., \& Ichimaru, S. 1983, ApJ, 273, 774

\noindent 
King, I. R., Anderson, J., Cool, A. M., \& Piotto, G. 1998, ApJ, 492, L37


\noindent 
Kolb, U., \& Baraffe, I. 1999, MNRAS, 309, 1034

\noindent 
Kolb, U., Rappaport, S., Schenker, K., \& Howell, S. 2001, ApJ, in
press (\mbox{astro-ph/0108322})


\noindent 
Marsh, T. R., et al.\ 2001,
in The Physics of Cataclysmic Variables and Related Objects,
ed.\ B. T. G\"ansicke, K.  Beuermann, \& K. Reinsch (San Francisco: ASP),
in press
(\mbox{astro-ph/0108334})

\noindent 
Menou, K. 2000, Science, 288, 2022

\noindent 
Nomoto, K. 1982, ApJ, 253, 798 

\noindent 
Osaki, Y. 1996, PASP, 108, 39 

\noindent 
Prialnik, D., \& Kovetz, A. 1995, ApJ,  445, 789

\noindent 
Pringle, J. E. 1988, MNRAS, 230, 587 

\noindent  
Shafter, A. W. 1992, ApJ, 394, 268

\noindent 
Shara, M. M., Livio, M., Moffat, A. F. J., \& Orio, M. 1986, ApJ, 311,
163 

\noindent 
Sion, E. M. 1985, ApJ, 297, 538 

\noindent 
Sion, E. M. 1995, ApJ, 438, 876

\noindent 
Sion, E. M. 1999, PASP, 111, 532


\noindent 
Szkody, P., Sion, E. M., G\"ansicke, B. T., \& Howell, S. B. \ 2001,
in The Physics of Cataclysmic Variables and Related Objects,
ed.\ B. T. G\"ansicke, K.  Beuermann, \& K. Reinsch (San Francisco: ASP),
in press

\noindent 
Szkody, P., et al.\ 2002, AJ, in press \mbox{(astro-ph/0110291)}

\noindent 
Taylor, J. M., Grindlay, J. E., Edmonds, P. D., \& Cool, A. M. 
2001, ApJ, 553, L169

\noindent
Townsley, D. M., \& Bildsten, L. 2001,
in The Physics of Cataclysmic Variables and Related Objects,
ed.\ B. T. G\"ansicke, K.  Beuermann, \& K. Reinsch (San Francisco: ASP),
in press

\noindent 
Urban, J., et al.\ 2000, PASP, 112, 1611



\noindent 
Wood, J. H. \& Crawford, C. S. 1986, MNRAS, 222, 645

\end{references}
\end{document}